\documentclass[twocolumn,showpacs,preprintnumbers,amsmath,amssymb,superscriptaddress]{revtex4}
\usepackage{graphicx}% Include figure files
\usepackage{bm}% bold math

\begin{document}
\title{Single-Dirac-cone Z$_2$ topological insulator phases in distorted Li$_2$AgSb-class and related quantum critical Li-based spin-orbit compounds}
%\title{Distorted Li$_2$M$'$X (M$'$=Cu, Ag, and Au, X=Sb and Bi): A new platform for topological insulators}% Force line breaks with \\
%\title{A new platform for topological quantum phenomena: Topological Insulator states in thermoelectric Heusler-related ternary compounds }% Force line breaks with \\

\author{H. Lin}
%\affiliation{Joseph Henry Laboratories of Physics, Princeton University, Princeton, New Jersey 08544, USA}
\affiliation{Department of Physics, Northeastern University, Boston, Massachusetts 02115, USA}

\author{L.A. Wray}
\affiliation{Joseph Henry Laboratories of Physics, Princeton University, Princeton, New Jersey 08544, USA}
\author{Y. Xia}
\affiliation{Joseph Henry Laboratories of Physics, Princeton University, Princeton, New Jersey 08544, USA}
\author{S.-Y. Xu}
\affiliation{Joseph Henry Laboratories of Physics, Princeton University, Princeton, New Jersey 08544, USA}
\author{S. Jia}
\affiliation{Department of Chemistry, Princeton University, Princeton, New Jersey 08544, USA,}
\author{R.J. Cava}
\affiliation{Department of Chemistry, Princeton University, Princeton, New Jersey 08544, USA,}
\author{A. Bansil}
\affiliation{Department of Physics, Northeastern University, Boston, Massachusetts 02115, USA}
\author{M.Z. Hasan}
\affiliation{Joseph Henry Laboratories of Physics, Princeton University, Princeton, New Jersey 08544, USA}
\affiliation{Princeton Center for Complex Materials, Princeton University, Princeton, New Jersey 08544, USA}
\affiliation{Princeton Institute for Science and Technology of Advanced Materials, PRISM, Princeton University, Princeton, New Jersey 08544, USA}
\email {mzhasan@Princeton.edu}

\date{April 6, 2010}% It is always \today, today,

\begin{abstract}
We have extended our new materials class search for the experimental realization of Z$_2$ topological insulators from binary [\textbf{Bi$_2$Se$_3$ class}, Xia \emph{et.al.}, Nature Phys.\textbf{5}, 398 (2009)] and the ternary [\textbf{Half-Heusler class}, Lin \emph{et.al.}, arXiv:1003.0155 (2010)] series to non-Heusler \textbf{Li-based ternary intermetallic series} Li$_2M'X$ ($M'$=Cu, Ag, and Au, $X$=Sb and Bi) with CuHg$_2$Ti-type structure. We discovered that the distorted-Li$_2$AgSb is a lightweight compound harboring a 3D topological insulator state with Z$_2$=-1 although the groundstate lies near a critical point, whereas the related Li$_2$CuSb-type compounds are topologically trivial. Non-Heusler ternary Li$_2M'X$ series (with a number of variant compounds) we identified here is a new platform for deriving novel stoichiometric compounds, artificial quantum-well/heterostructures, nano-wires, nano-ribbons and nanocrystals. We have grown some of these bulk materials (experimental results will be reported separately).
\end{abstract}

\maketitle

Topological insulators (TI)\cite{review, Kane1st,FuKane,DavidNat1,KonigSci,DavidTunable,TIbasic,BernevigSciHgTe,Roy,MooreandBal} realize a novel state of quantum matter that are distinguished by topological invariants of bulk band structure rather than spontaneously broken symmetries. Its material realization in 2D artificial HgTe-quantum wells \cite{KonigSci} and 3D Bi-based binary compounds \cite{DavidNat1,DavidTunable,DavidScience,MatthewNatPhys, Noh, ZhangPred,ChenBiTe,BiTeSbTe, Roushan, WrayCuBiSe} led to a surge of interest in discovering novel topological physics in world-wide condensed matter community. A number of exotic quantum phenomena have been predicted to exist in multiply-connected geometries \cite{Majorana, ZhangDyon, KaneSCproximity, KaneDevice, cenke, palee, dhlee} which require an enormous amount of materials flexibility. Given the right materials, these topological properties naturally open a window to new realms of spintronics and quantum computing. Just as the majority of normal metals are neither good superconductors nor strongly magnetic, there is no inherent likelihood that the band structure of topological insulators known so far will be suitable for the realization of any of the novel physical states or devices that have been predicted for quantum computing or spintronics. We need to expand our search.
We have extended our previous search for TI materials from binary (Bi$_2$X$_3$ series, Xia \emph{et.al.}, Nature Phys. \textbf{5}, 398 (2009)) and the thermoelectric ternary compounds (half-Heuslers, Lin \emph{et.al.}, arXiv:1003.0155\cite{heuslerhasan}) to Li$_2M'X$ ($M'$=Cu, Ag, and Au, $X$=Sb and Bi) series. We discover that the distorted Li$_2$AgSb is the best lightweight ternary compound of the ``$M_2M'X$" series harboring a 3D topological insulator state with Z$_2$=-1. We also show that the Li$_2M'X$ series is a natural platform that hosts a range of candidate compounds, alloys and artificial heterostructures (quantum-wells). We also discovered several different paradigms of trivial and non-trivial topological ordering in this class, including an intrinsically metallic nontrivial topological state in Li$_2$AgBi and Li$_2$AuBi.
In this Letter, we use first principles theoretical calculations to show that distorted Li$_2$AgSb is the best lightweight ternary compound expected to realize a three dimensional topological insulator state.

The crystal lattice of Li$_2M'X$ ($M'$=Cu, Ag, and Au, $X$=Sb and Bi) is described by the space group
$F\bar{4}3m$, with the atomic arrangement presented in Fig. 1A.
$M'$ and $X$ atoms occupy the Wyckoff 4d and 4a positions, respectively. Li atoms fill the remaining empty space
in 4b and 4c positions. Because $M'$ and $X$ atoms form a zincblende type sublattice,
these materials resemble the topologically nontrivial gray tin. There is no spatial inversion symmetry in Li$_2M'X$ compounds. These observations suggest that Li$_2M'X$ could be candidates for 3D Z$_2$ topological insulators if some odd number of band inversions are realized.

%The crystal lattice of Li$_2$AgSb is described by the space group $F\bar{4}3m$,
%with the atomic arrangement shown in Fig.~\ref{fig:sketch}a.
%Crystalline compounds of related compounds are assigned the chemical formula ``Li$_2M'X$", where $M'$ and $X$ atoms
%occupy the Wyckoff 4d and 4a positions respectively \cite{Wyckoff}. The $M'$ and $X$ atoms form a zincblende lattice,
%when taken alone. Li fill empty space at Wyckoff 4b and 4c within the zincblende structure.
%These materials closely resemble the previously known topologically nontrivial compound
%HgTe \cite{BernevigSciHgTe,KonigSci}, which achieves topological order with a zincblende structure.
%There is no spatial inversion symmetry in zincblende and Li$_2$AgSb.

%Assume the nuclear charge $Z$ of the atoms at
%the Li, Ag, and Sb positions are $Z_{M}=3-0.5 x-0.5 y$, $Z_{M'}=47+x$, and $Z_X=51+y$, repectively.
%The continuous mapping can start with
%x=0 and y=0 corresponds to Li$_2$AgSb, and end with x=1 and y = 1 corresponds
%to an artificial compound He$_2$CdTe. Since hellium is chemical inert, we expect He$_2$CdTe have similar
%band strucutres as CdTe.
% These observations suggest that
%Li$_2$AgX could be candidates for 3D Z2 topological insulators if some odd
%number of band inversions are realized within the crystal Brillouin zone induced by the
%spin-orbit coupling.

\begin{figure*}
\includegraphics[width=15cm]{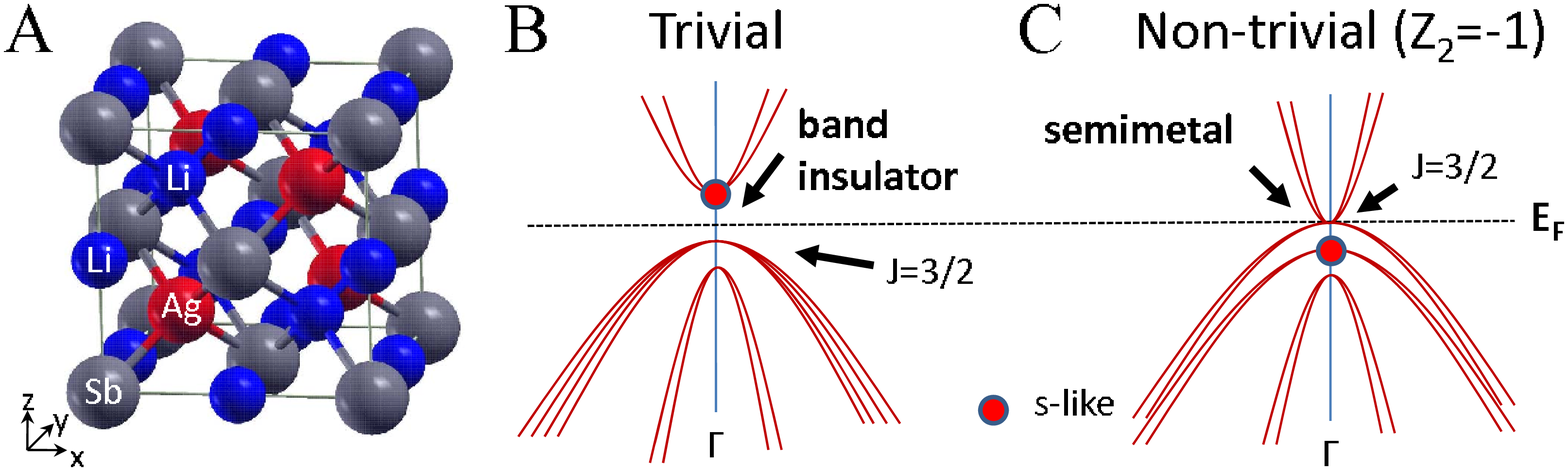}
\caption{\label{fig:sketch}
\textbf{Li$_2$AgSb crystal structure and topological band inversion}: \textbf{a} The crystal structure of Li$_2$AgSb. Li, Ag, and Sb are denoted by blue, red, and gray balls. Sb and Ag form the zinc-blende structure.
Diagrams in \textbf{b}-\textbf{c} illustrate band structures near the $\Gamma$-point for trivial and non-trivial cases, respectively. Red dots denote the s-like orbitals at $\Gamma$-point. Band inversion occurs in the non-trival case where the s-like orbitals at the $\Gamma$-point are below the four-fold degenerate $j=3/2$ bands.
}
\end{figure*}

%Hg$_{1-x}$Cd$_x$Te is a family of strong spin orbit 3D materials (Dornhaus and Numtz, 1983)\cite{HgTe}. In the electronic structure of CdTe the conduction band edge states have an \emph{s} like symmetry, while the valence band edge states have a \emph{p} like symmetry. In HgTe, the \emph{p} levels rise above the \emph{s} levels, leading to an inverted band structure. Due to this similarity, it is instructive to begin our discussion of Li$_2M'X$ band structure topology by examining the nontrivial topological nature of well studied 3D-HgTe. While the related material, CdTe
%(half-Heusler []CdTe)
%is a typical band insulator, HgTe is topologically nontrivial due to the inversion of two groups of bands with respective \emph{s}- and \emph{p}- orbital symmetry at the $\Gamma$-point. The different ordering of band symmetries at the $\Gamma$-point for trivial and non-trivial cases are summarized in Fig.~\ref{fig:sketch}b-c, with red dots labeling the bands with large \emph{s}-orbital occupancy. In the trivial case, the \emph{s}-like orbital is above the band gap, and in the nontrivial case, the \emph{s}-like orbital at the $\Gamma$-point is below the four-fold degenerate $j=3/2$ bands \cite{BernevigSciHgTe,FuKane}. This band inversion only occurs near the $\Gamma$-point and is absent at other special time reversal invariant points. We will show that Li$_2M'X$ compounds also exhibit these two topologically distinct classes, distinguished in the same way through orbital symmetries on the $X$-sublattice.

First-principles band calculations were performed with the linear augmented-plane-wave (LAPW) method using the WIEN2K package \cite{wien2k} in the framework of density functional theory (DFT). The generalized gradient approximation (GGA) of Perdew, Burke, and Ernzerhof \cite{PBE96} was used to describe the exchange-correlation potential. Spin orbital coupling (SOC) was included as a second variational step using a basis of scalar-relativistic eigenfunctions. The calculated (DFT-GGA) band structures of Li$_2$AgSb, Li$_2$AgBi, Li$_2$AuBi, and Li$_2$CuSb along high symmetry lines in the Brillouin zone are presented in Fig.~\ref{fig:bulkbands}. A common feature of these materials is that the top of the valence band is located at the $\Gamma$-point. For Li$_2$AgSb, away from $\Gamma$, the Fermi level is completely gapped. Therefore, the topological properties can be determined from observations of band structure only near the $\Gamma$-point. For Li$_2$AgBi and Li$_2$AuBi, part of the conduction band along L-$\Gamma$ is below the $E_F$ and forms electron pockets near the L-point. The conduction band and valence band overlap and the topological index Z$_2$ is not well defined. If the overlap is removed by distortion, Z$_2$ can also be determined from the band structure only near the $\Gamma$-point.
%Other materials with similar properties are addressed in the supplementary information.

Confining our view to band structure very close to the Fermi level (Fig.~\ref{fig:bulkbands}a(inset)), we find that the orbital angular momentum symmetries of Li$_2M'X$ compounds are identical to those defining low energy properties in HgTe and CdTe. For Li$_2$AgBi and Li$_2$AuBi, two upward-concave bands and two downward-concave bands are degenerate at the $\Gamma$ point.
%The Fermi energy should lie exactly at this degenerate point for undoped samples.
These four-fold degenerate states at the $\Gamma$-point have \emph{p}-type orbital symmetry with a total angular momentum eigenvalue of $j=3/2$, and lie above a two-fold degenerate s-like state, representing an inversion relative to the natural order of \emph{s}- and \emph{p}-type orbital derived band structure. Away from the $\Gamma$-point, the upward dispersing bands gain significant \emph{s}-like character due to orbital hybridization.
Thus, by analogy with HgTe, we expect Li$_2$AgBi and Li$_2$AuBi to be a topologically nontrivial metal (or ``topological metal"), as is the case with elemental antimony \cite{DavidScience}, and its conductivity could probably be manipulated through alloying, just as antimony was alloyed with bismuth to discover the first known example of a three dimensional TI \cite{DavidNat1} (Bi$_{1-x}$Sb$_{x}$). The s-type bands in Li$_2$CuSb are above $E_F$ and above the $j=3/2$ bands, meaning that its band structure lacks the $s/p$ inversion that leads to strong topological order.
For Li$_2$AgSb, the s-type and p-type states at the $\Gamma$-point is nearly degenerate. The conduction band and one of the valence band have almost linear dispersion around the $\Gamma$-point, indicating it is on the critical point of the topological phase. With a small expansion of lattice, the $s/p$ band inversion occurs as shown in Fig. 3(b).
As we will establish more rigorously below, these symmetries correctly indicate that Li$_2$CuSb is topologically trivial while expanded Li$_2$AgSb is topologically nontrivial.

\begin{figure*}
\includegraphics[width=15cm]{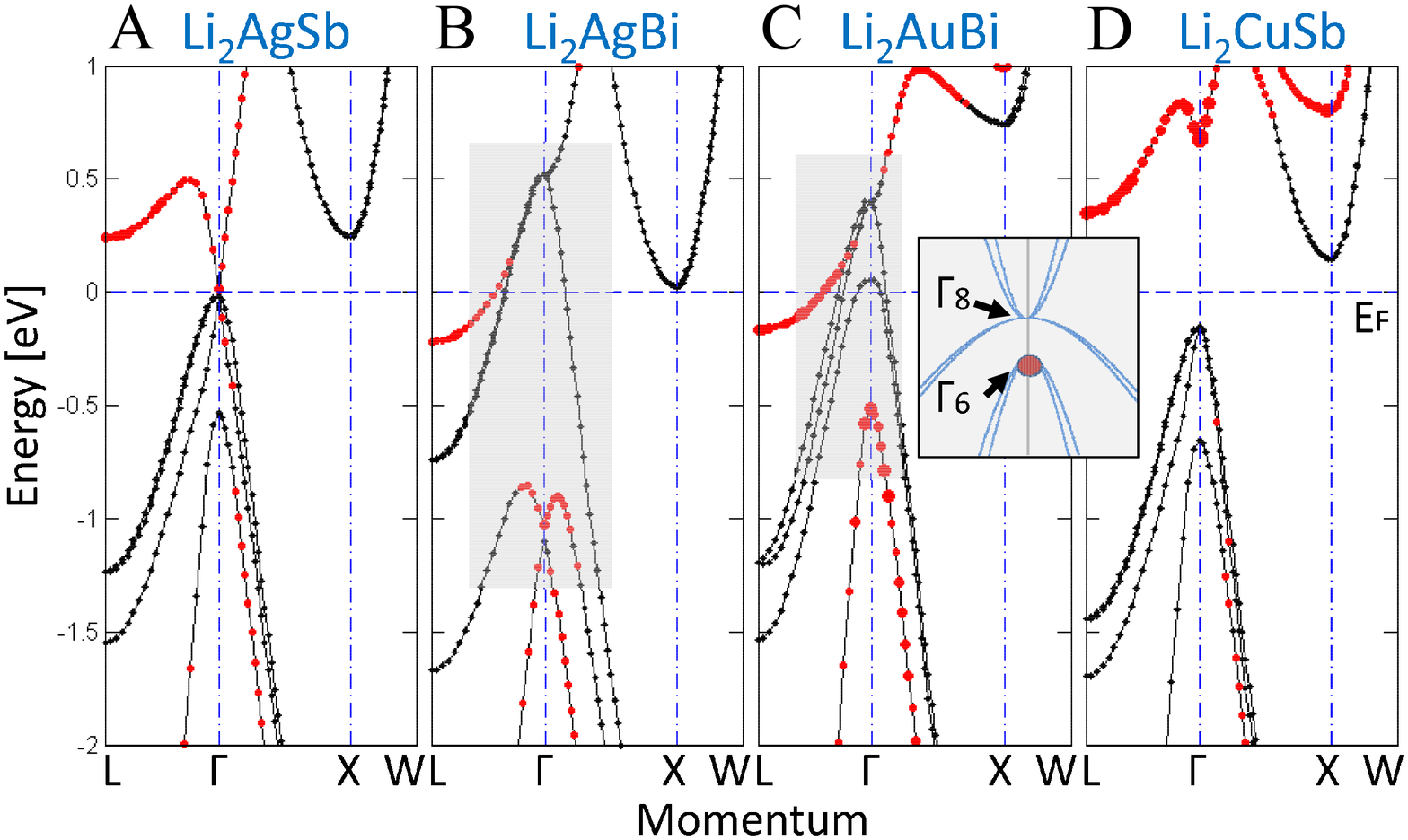}
\caption{\label{fig:bulkbands}
\textbf{Li$_2M'X$ electronic structures}: Bulk band structures of (\textbf{a}) Li$_2$AgSb, (\textbf{b}) Li$_2$AgBi, (\textbf{c}) Li$_2$AuBi, and (\textbf{d}) Li$_2$CuSb. The size of red data points is proportional to the probability of s-orbital occupation on the anion atom (``$X$"). An inset in panel \textbf{c} highlights inverted band symmetries associated with topological order, corresponding to the the diagram in Fig. 1c.
}
\end{figure*}

In terms of the symmetry notation that has been applied to HgTe in previous literature, our calculations show that bands near $E_F$ at the $\Gamma$-point possess $\Gamma_6$ (2-fold degenerate), $\Gamma_7$ (2-fold degenerate), and $\Gamma_8$ (4-fold degenerate) symmetry in all of these compounds. Both expanded Li$_2$AgSb and HgTe have $\Gamma_8$ states at $E_F$. The $\Gamma_6$ symmetry bands are below $\Gamma_8$ and occupied, providing the band inversion that distinguishes topological order. Although the order of $\Gamma_7$ and $\Gamma_6$ is different, it is not relevant to the topological nature since both are occupied. In the band structures of topologically trivial band insulators Li$_2$CuSb and CdTe, the $\Gamma_6$ states are above the downward dispersing valence bands. These non-inverted $\Gamma_6$ states lie above $E_F$ and are unoccupied. It is due to the occupancy of $\Gamma_6$ bands at the $\Gamma$-point in Li$_2$AgSb that the $Z_2$ topological invariant picks up an extra factor of -1 when compared to Li$_2$CuSb.

\begin{figure*}
\includegraphics[width=15cm]{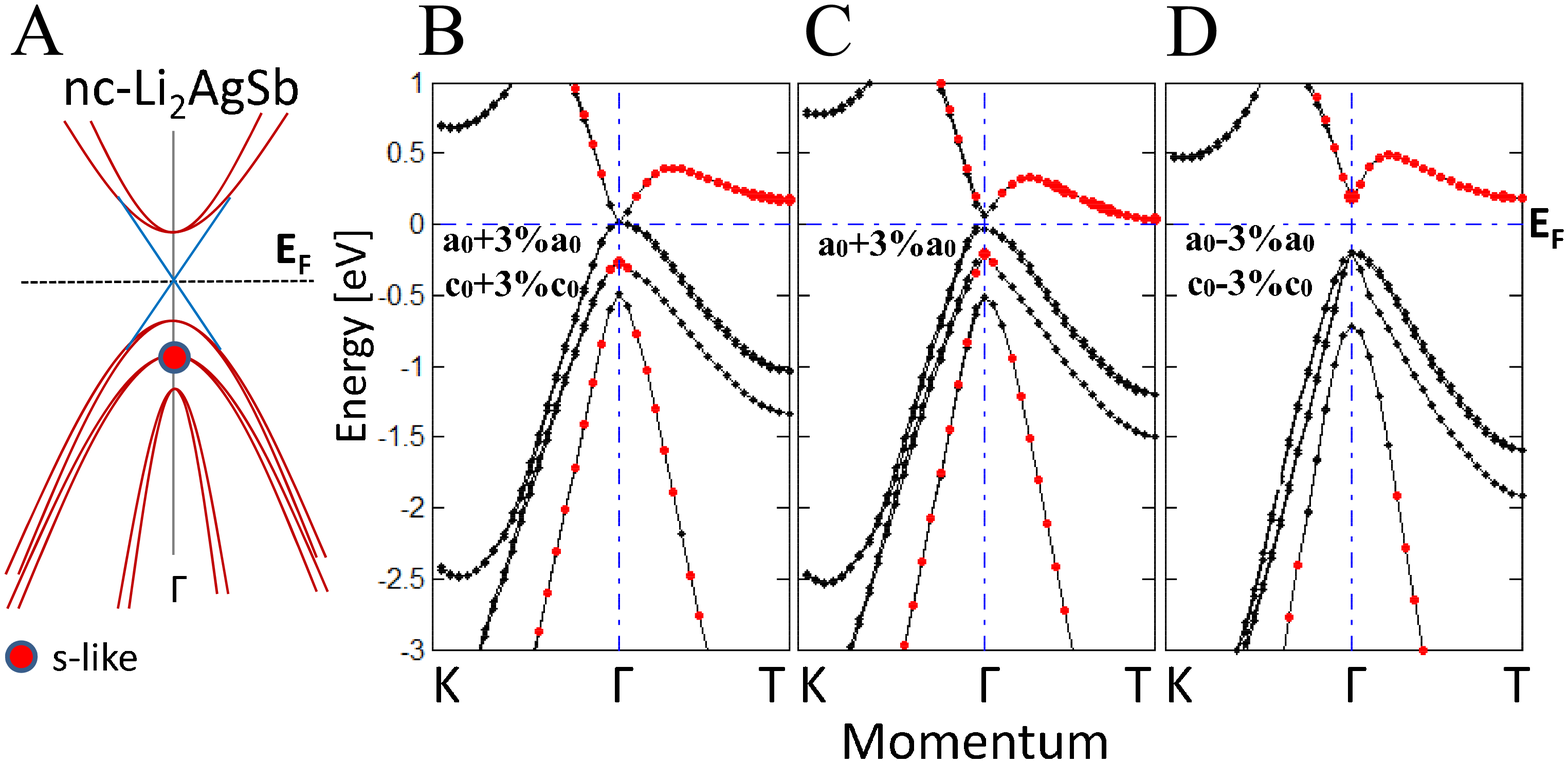}% Here is how to import EPS art
\caption{\label{fig:distortion}
\textbf{Topological insulator state with a single-Dirac-cone}: (a) A sketch of band structure near the $\Gamma$-point for topologically non-trivial Li$_2$AgSb with a lattice distortion. The s-like $\Gamma_6$ states are marked with a red dot. Lattice distortion causes a gap to open at $E_F$, resulting in a topological insulator state. For such a topological insulator, the surface bands will span the bulk band gap and should have an odd number of Dirac cones, resembling the dispersion plotted with blue lines. (b) Band structures of expanded Li$_2$AgSb where $a=a_0+3\%a_0$ and $c=c_0+3\%c_0$, where $a_0$ and $c_0$ correspond to the experimental lattice. (b) Band structures of rhombohedral distorted Li$_2$AgSb with lattice constants $a=a_0+3\%a_0$ and $c=c_0$  (d) Band structures of Li$_2$AgSb under hydrostatic pressure where $a=a_0-3\%a_0$ and $c=c_0-3\%c_0$.
}
\end{figure*}

The fact that none of the topologically nontrivial materials discussed in this work (3D-HgTe, expanded Li$_2$AgSb, Li$_2$AgBi, and Li$_2$AuBi) are naturally insulating known to be due to the $\Gamma$-point degeneracy of positive- and negative-mass bands with $\Gamma_8$ symmetry, which results from the crystal symmetry group. Distortion of the crystal lattice through static pressure or finite size effects can lift the degeneracy and open a gap at $E_F$, causing a direct gap between the topologically inverted valence and conduction bands and resulting in the appearance of topologically defined surface states. Fu and Kane have previously discussed this possibility for 3D-HgTe\cite{FuKane}. In Fig.~\ref{fig:bulkbands}c, we demonstrate that a perturbative rhombohedral lattice distortion caused by uniaxial pressure will lift the band structure degeneracy in Li$_2$AgSb, resulting in a fully gapped topological insulator state. With reference to the convention cubic structure, a 3\% expansion is applied in the plane perpendicular to [111]. As has been observed when distortion is applied to the zincblende structured binary compound HgTe, bulk band structure of the distorted Li$_2$AgSb is fully gapped and realizes strong Z$_2$ topological order.

Conversely, the topological band inversion in Li$_2$AgSb can be removed altogether by uniformly decreasing all lattice constants, demonstrating the sensitive chemical tunability in this TI class (Fig.~\ref{fig:bulkbands}d). After a 3\% reduction in all lattice parameters, the s-like $\Gamma_6$ symmetry bands are observed to rise above the $\Gamma_8$ bands, removing band inversion at the $\Gamma$-point and resulting in a topologically trivial band insulator state.

In conclusion, we have discovered a new class of topological insulator materials realized by distorted Li$_2M'X$, and demonstrated band structure topology calculations for representative topologically critical semimetal (Li$_2$SbAg) and nontrivial metallic (Li$_2$AgBi and Li$_2$AuBi) materials. Bulk band structure in distorted Li$_2$SbAg is shown to be characterized by Z$_2$=-1.
%A large number of topological compounds exist (list to be discussed elsewhere) in this ternary material class, making it
%a versatile platform for exploring many different device configurations for realizing topological quantum phenomena not
%accessible in the binary topological insulator classes such as the Bi$_2$Se$_3$ series discovered previously.

%M.Z.H. acknowledges discussions with C.L. Kane and B.A. Bernevig and support from U.S.DOE and A.P. Sloan Research Fellowship. H.L. acknowledges support from Northeastern and Princeton University. R.J.C. acknowledges discussions and long-standing collaborations with C. Felser and T. Kilmczuk on thermoelectric and superconducting-Heusler phases and with Y.S. Hor on Ternary-topological-materials.

%\bibliography{draft}% Produces the bibliography via BibTeX.

\begin{figure*}
\includegraphics[width=13cm]{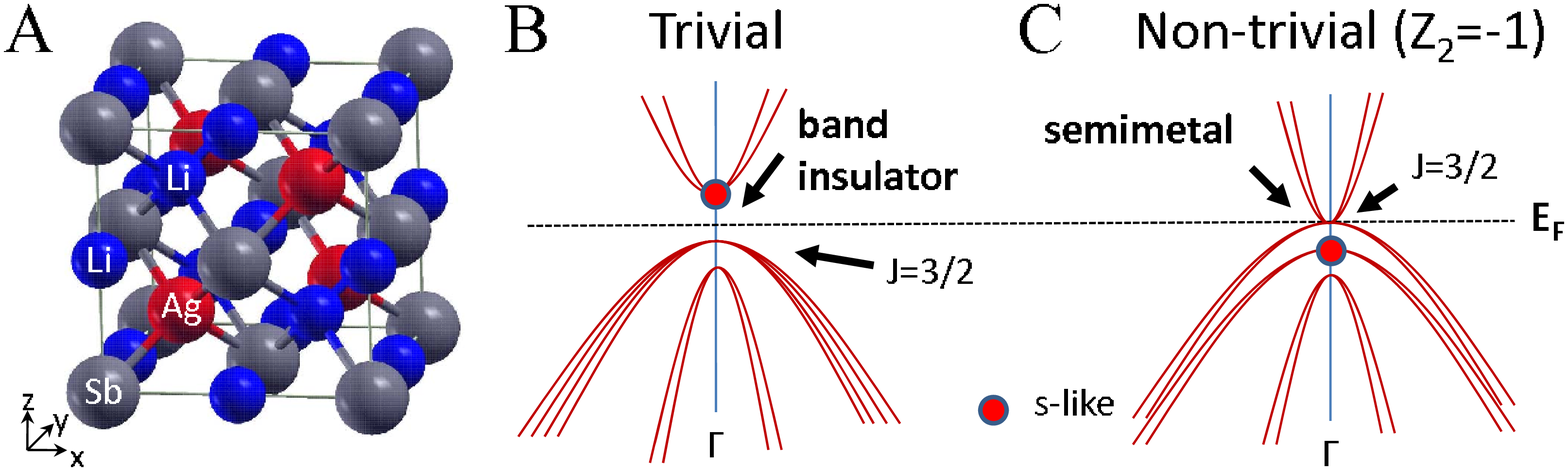}
\caption{\label{fig:sketch}
\textbf{Li$_2$AgSb crystal structure and band inversion}: \textbf{a} The crystal structure of Li$_2$AgSb. Li, Ag, and Sb are denoted by blue, red, and gray balls. Sb and Ag form the zinc-blende structure.
Diagrams in \textbf{b}-\textbf{c} illustrate band structures near the $\Gamma$-point for trivial and non-trivial cases, respectively. Red dots denote the s-like orbitals at $\Gamma$-point. Band inversion occurs in the non-trival case where the s-like orbitals at the $\Gamma$-point are below the four-fold degenerate $j=3/2$ bands.
}
\end{figure*}

\begin{figure*}
\includegraphics[width=13cm]{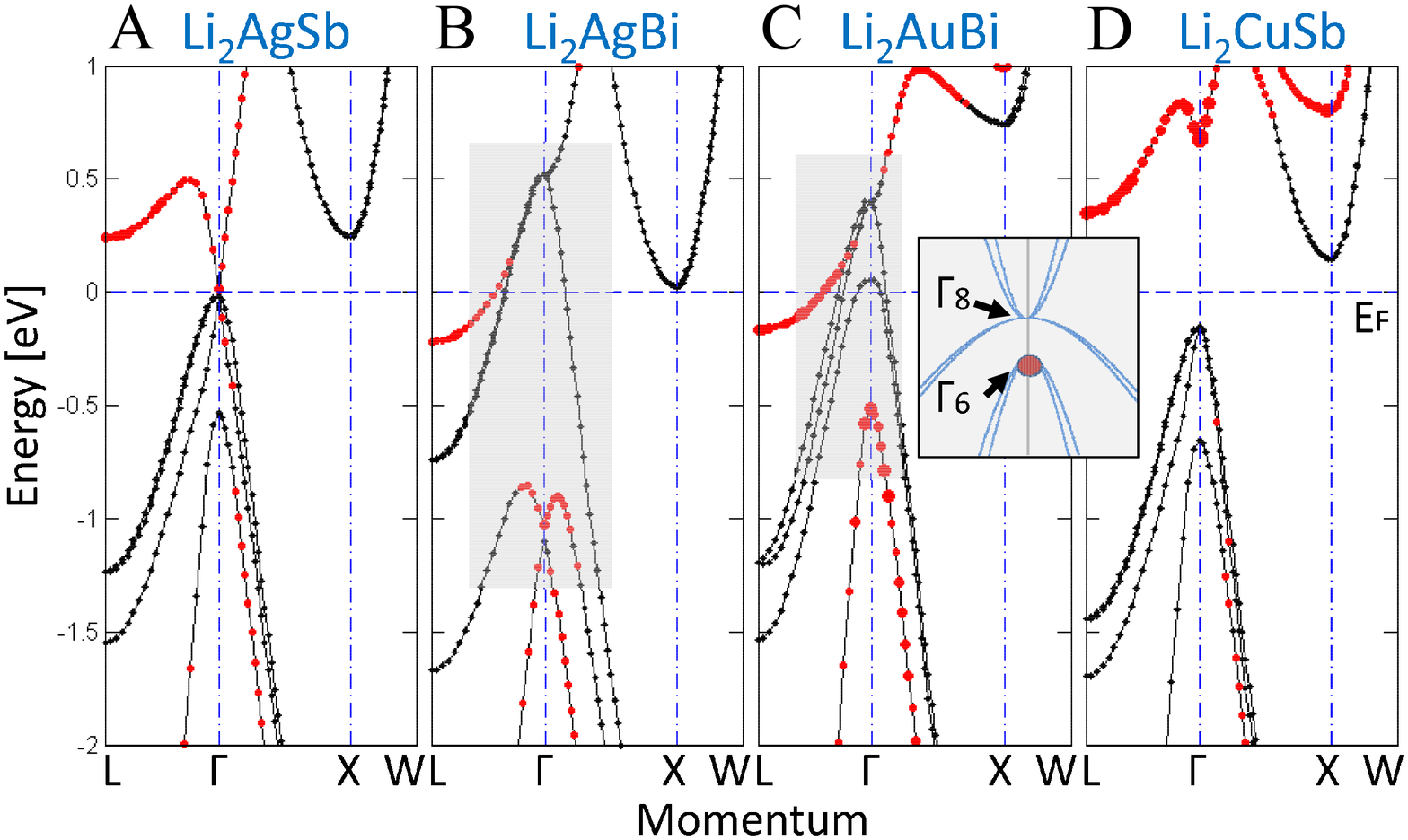}
\caption{\label{fig:bulkbands}
\textbf{Li$_2M'X$ band structures}: Bulk band structures of (\textbf{a}) Li$_2$AgSb, (\textbf{b}) Li$_2$AgBi, (\textbf{c}) Li$_2$AuBi, and (\textbf{d}) Li$_2$CuSb. The size of red data points is proportional to the probability of s-orbital occupation on the anion atom (``$X$"). An inset in panel \textbf{c} highlights inverted band symmetries associated with topological order, corresponding to the the diagram in Fig. 1c.
}
\end{figure*}

\begin{figure*}
\includegraphics[width=12cm]{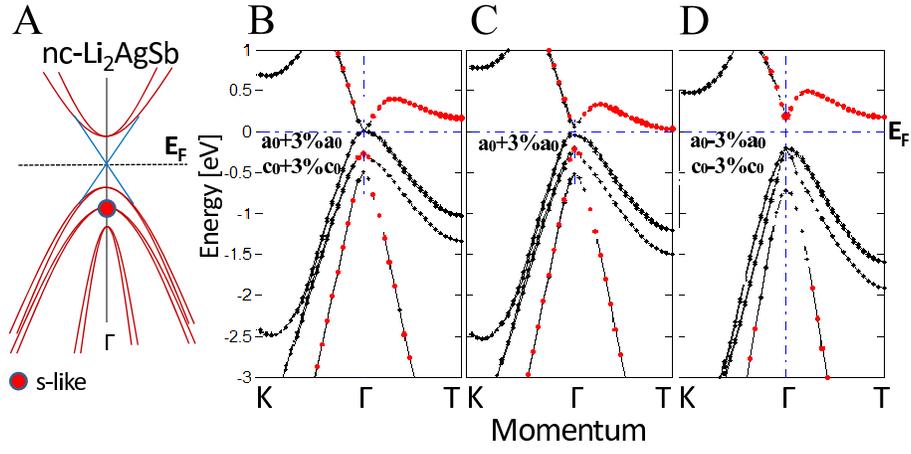}% Here is how to import EPS art
\caption{\label{fig:distortion}
\textbf{The topological insulator state with a single-Dirac-cone}: (a) A sketch of band structure near the $\Gamma$-point for topologically non-trivial Li$_2$AgSb with a lattice distortion. The s-like $\Gamma_6$ states are marked with a red dot. Lattice distortion causes a gap to open at $E_F$, resulting in a topological insulator state. For such a topological insulator, the surface bands will span the bulk band gap and should have an odd number of Dirac cones, resembling the dispersion plotted with blue lines. (b) Band structures of expanded Li$_2$AgSb where $a=a_0+3\%a_0$ and $c=c_0+3\%c_0$, where $a_0$ and $c_0$ correspond to the experimental lattice. (b) Band structures of rhombohedral distorted Li$_2$AgSb with lattice constants $a=a_0+3\%a_0$ and $c=c_0$  (d) Band structures of Li$_2$AgSb under hydrostatic pressure where $a=a_0-3\%a_0$ and $c=c_0-3\%c_0$.
}
\end{figure*}

\end{document}